\documentclass[
reprint, 
 amsmath,amssymb,
aps, 
]{revtex4-2}

\usepackage{hyperref}
\usepackage{graphicx}
\usepackage{dcolumn}
\usepackage{bm}

\graphicspath{{./}{./Figures/}}

\usepackage{color, soul}
\definecolor{lightyellow}{RGB}{255, 255, 204}
\definecolor{lightblue}{RGB}{204, 238, 255}

\usepackage{xspace}
\newcommand{\ecoli}{\emph{E.~coli}\xspace}

\newcommand{\avg}[1]{\mathbb{E}[#1]}
\newcommand{\var}[1]{\text{Var}[#1]}
\newcommand{\covar}[1]{\text{Cov}[#1]}
\newcommand{\cv}[1]{\text{CV}_{#1}}

\begin{document}


\title{\textbf{Essential Role of Extrinsic Noise in Models of \ecoli Division Control}}

\author{Mattia Corigliano \textsuperscript{1, 2}}
\thanks{These authors contributed equally to this work}
\author{Kuheli Biswas \textsuperscript{4}}%
\thanks{These authors contributed equally to this work}
\author{Matteo Bocchiola \textsuperscript{2}}%
\author{Daniele Montagnani \textsuperscript{2}}%
\thanks{Current affiliation: Dipartimento di Matematica, Università di Pavia, Via Ferrata 5, 27100, Pavia}
\author{Ariel Amir \textsuperscript{4}}
\thanks{Correspondence: ariel.amir@weizmann.ac.il, marco.cosentino-lagomarsino@ifom.eu}
\author{Marco Cosentino Lagomarsino \textsuperscript{1, 2, 3}}
\thanks{Correspondence: ariel.amir@weizmann.ac.il, marco.cosentino-lagomarsino@ifom.eu}
 \affiliation{%
 1. IFOM-ETS, The AIRC Institute of Molecular Oncology, via Adamello 16, 20139 Milan, Italy
}
\affiliation{%
2. Dipartimento di Fisica, Università degli studi di Milano, via Celoria 16, 20133, Milan, Italy
}
\affiliation{%
3. Istituto Nazionale di Fisica Nucleare, Sezione di Milano, via Celoria 16, 20133, Milan, Italy
}
\affiliation{
4. Department of Physics of Complex Systems, Weizmann Institute of Science, Rehovot 7610001, Israel
}


\begin{abstract}
  Our understanding of cell division control in bacteria still relies largely on
  interpreting correlations between phenomenological variables, with
  limited connection to the underlying molecular mechanisms.
  Here, we analytically solve a stochastic threshold–accumulation
  model in which a size-dependent divisor protein triggers division
  upon reaching a noisy, autocorrelated threshold, quantifying within
  a unified framework the combined effects of intrinsic and extrinsic
  noise and key mechanistic parameters such as protein reset and
  threshold memory.
  We show that incorporating these elements yields behavior far richer
  than the commonly assumed \emph{adder}, spanning a continuum of
  division strategies from \emph{timer} to \emph{sizer} while
  modulating size fluctuations in a nontrivial fashion.  Comparison
  with single-cell \ecoli data shows that extrinsic noise and
  additional mechanistic ingredients are required to account for the
  observed size fluctuations. The \emph{adder} emerges when threshold
  correlations balance protein reset, generalizing the hypothesis that
  full reset is necessary to maintain \emph{adder} control.
  Our results establish a unified analytical framework linking stochastic molecular processes to emergent division laws, to be used in more complex bacterial cell-cycle models.
\end{abstract}

\maketitle


Cell-cycle progression requires maintaining key physiological
variables, including cell size, within specific functional
ranges~\cite{Ginzberg2015,Neurohr2019,Neurohr2020,Xie2022,Lengefeld2023}. Single
\ecoli\ cells, for example, divide on average after adding a constant
size~\cite{Campos2014,Taheri-Araghi2015}. This division
strategy, known as the \emph{adder}, ensures a stable birth-size
distribution with fluctuations of only $10$--$20\%$ around the
mean. Similar \emph{adder}-like behavior and comparable size fluctuations
have been observed across diverse organisms, including budding
yeast~\cite{Soifer2016,Chandler-Brown2017,Garmendia-Torres2018},
archaea~\cite{Eun2017}, and mammalian cells~\cite{Cadart2018, Levien2026}, leading
to questions regarding its mechanistic origin(s).

Yet our understanding remains largely phenomenological, often relying on linearized frameworks linking fluctuations in birth size, added size, division time, and growth rate~\cite{Amir2014,Grilli2017,Ho2018,Grilli2018}. These analytical
approaches allow us to characterize stable size
distributions~\cite{Ho2018} and to test competing models of division
control~\cite{Micali2018a,Micali2018b,Colin2021,Tiruvadi-Krishnan2022,Kar2023},
but cannot explain how the observed division-control laws and size
fluctuations arise from underlying molecular control processes and
molecular noise sources.

To address this gap, we analyze a class of models in which a divisor
molecule is synthesized at a rate proportional to cell size and
triggers division upon reaching a threshold copy
number~\cite{Harris2016, Barber2017, Ojkic2019, Si2019, Pandey2020, Panlilio2021}. These
``threshold–accumulation'' models have been proposed as minimal,
biologically grounded implementations of the \emph{adder} strategy in \ecoli~\cite{Harris2016, Barber2017, Si2019, Pandey2020, Panlilio2021}

Here, we develop an analytical framework that incorporates stochastic
protein accumulation (leading to intrinsic first-passage time noise),
threshold variability (extrinsic noise), and key mechanistic
ingredients such as molecule reset and threshold memory, enabling a
unified treatment of intrinsic and extrinsic sources of
variability. This approach bridges and extends previous studies that
considered intrinsic fluctuations~\cite{ElGamel2024} or stochastic
thresholds~\cite{Luo2023} separately, and directly links division
control and size fluctuations to underlying molecular parameters.

We consider exponentially growing cells undergoing successive
divisions at constant growth rate $\alpha$. The empirically observed
stability of the size distribution implies the presence of active
division control~\cite{Amir2014}. Phenomenologically, this can be
described as a coupling between fluctuations in the logarithmic birth
size $q_0^{(i)}\equiv \log(s_0^{(i)}/s^*)$ and in the logarithmic
added size $G^{(i)}\equiv \log(s_f^{(i)}/s_0^{(i)})=\alpha\tau^{(i)}$,
where $s_0^{(i)}$ and $s_f^{(i)}$ denote the sizes at birth and
division, and $\tau^{(i)}$ the division time in cycle $i$. Exploiting
the smallness of size fluctuations, this relation is typically
linearized~\cite{Amir2014,Grilli2017} as
$ \delta G^{(i)} = -\lambda_G \delta q_0^{(i)} + \eta_G^{(i)}, $ with
$0\le\lambda_G\le1$. In this regime the logarithmic birth size follows
a discrete-time Ornstein–Uhlenbeck process and the single parameter
$\lambda_G$ encapsulates the effective division strategy. The
canonical strategies correspond to $\lambda_G=0.5$ (\emph{adder}: division
after a constant added size), $\lambda_G=1$ (\emph{sizer}: division at
a target size), and $\lambda_G=0$ (\emph{timer}: division after a
fixed time). Because $\lambda_G$ is directly linked to measurable
statistics in single-cell data~\cite{Amir2014,Grilli2017}, this
framework enables inference of division strategies and tests of
competing models~\cite{Micali2018b}. However, it remains
phenomenological and does not specify the underlying molecular
dynamics.

\begin{figure}[htpb] \centering \includegraphics[width=0.47\textwidth]{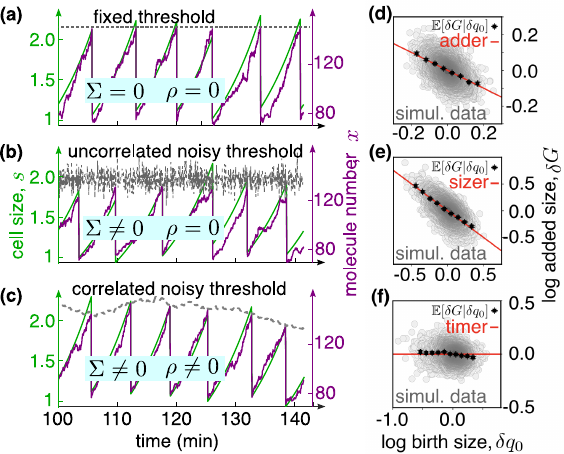}
    \caption{\textbf{Model dynamics and threshold regimes.}
      Exponentially growing single cells undergoing successive
      divisions. Division occurs when a divisor molecule $x(t)$,
      produced proportionally to cell size $s(t)$, reaches a threshold
      that is either (a) constant or (b,c) fluctuating, with
      uncorrelated or autocorrelated dynamics and memory $\rho$. The
      noise ratio is $\Sigma = CV_{\mathrm{th}}/CV_{\mathrm{mol}}$,
      with $\Sigma=0$ in (a) and $\Sigma>0$ in (b,c).}
    \label{fig:1}
\end{figure}

We therefore introduce a stochastic threshold–accumulation model
defined by the following ingredients. During cell cycle $(i)$, a
divisor molecule, ``$x$'', is synthesized at rate, $\mu s^{(i)}$,
proportional to cell size, and division occurs when its copy number,
$x^{(i)}$, first reaches a threshold
$\theta^{(i)}$~\cite{Harris2016,Barber2017, Ojkic2019, Si2019, Pandey2020, Panlilio2021}
[Fig.~\ref{fig:1}]. At division, cell size is halved and the molecule
is reset to $x_0^{(i+1)} = r^{(i)} \theta^{(i)}$. In what follows, we
consider a constant reset, $r$, set either to $r=0$ (full reset) or
$r=1/2$ (half reset), leaving the general case to the SI. Assuming
deterministic exponential growth and neglecting molecular and
threshold fluctuations, one has \begin{equation} s_f^{(i)} =
  s_0^{(i)}e^{\alpha \tau^{(i)}} = s_0^{(i)} + \tfrac{\alpha}{\mu}
  (1-r) \theta.
\label{eq:minimalTA}
\end{equation} This simple result has led to the widespread
interpretation of this model as a mechanistic realization of the
\emph{adder}~\cite{Harris2016, Panlilio2021}.  
We show below, however, that once stochastic effects are accounted
for, this correspondence holds only in restricted regions of parameter
space.

Accounting for molecular noise, division timing, $\tau^{(i)}$, becomes
the first-passage time of a non-homogeneous Poisson process,
$\lbrace x(t) \rbrace$, with rate, $\mu s_0^{(i)}e^{\alpha t}$, to a
fixed threshold, $\theta$ [Fig.~\ref{fig:1}(a)].  This intrinsic-noise
limit was recently analyzed in a related framework in which additional
degradation and constant production terms were introduced in the
synthesis dynamics of the divisor protein~\cite{ElGamel2024}. In the
absence of such terms, one has (see SI)
\begin{equation}
  \lambda^{\text{intr.}}_G=1/2; \quad
  (\mathrm{CV}^2_{s_0})^{\text{intr.}}=[3(1-r)\theta]^{-1},
  \label{eq:intrinsic-limit}
\end{equation}
implying \emph{adder} division control and Poisson scaling,
$\mathrm{CV}_{s_0}\propto 1/\sqrt{\theta}$, of size fluctuations. Reproducing the experimentally observed size variability
($\mathrm{CV}_{s_0}\approx 10$–$20\%$) would then require
$\theta\approx 8$–$70$ molecules~\cite{Pandey2020}—one to two orders of magnitude smaller than realistic copy numbers of candidate division proteins, including Fts proteins such as FtsZ and FtsN~\cite{Pla1991, Rueda2003, Ursinus2004, Mohammadi2009, Erickson2010, Li2014, Liu2015, Vischer2015, Egan2015}, as well as peptidoglycan precursors~\cite{Ramey1978, Van_Heijenoort1992, Van_Heijenoort2007}. Thus, within the purely size-proportional synthesis model studied here, intrinsic noise alone cannot account for the observed magnitude
of size fluctuations.  

One possible route to increase size variability is to introduce constant
(size-independent) production ($\nu$) or degradation ($\gamma$) terms
into protein synthesis dynamics, as explored in Ref.~\cite{ElGamel2024}. Each term increases birth-size
fluctuations and shifts division control away from the intrinsic
\emph{adder}: constant production biases toward timer-like behavior,
whereas degradation promotes sizer-like regimes (see SI). Within the broader
family of synthesis models parameterized by $(\mu,\nu,\gamma)$, the
\emph{adder} strategy was therefore identified in Ref.~\cite{ElGamel2024} as the
one minimizing cell-size noise at fixed threshold level. While
intriguing, whether such a minimization principle is biologically
realized remains unclear. More importantly, our work shows that even if
realized it does not uniquely support an \emph{adder} strategy, as noise
minimization preserves \emph{adder} behavior only for a constant threshold,
whereas threshold fluctuations and other mechanistic parameters shift
the minimum away from the \emph{adder} (e.g. see SI Fig.~1).

Here we follow a different route. We retain the experimentally supported
size-proportional synthesis law and instead introduce additional
variability at the level of the division threshold
[Fig.~\ref{fig:1}(b–c)]. Despite its minimal structure, the model remains analytically tractable and, as we
show below, rich enough to generate diverse behaviors
[Fig.~\ref{fig:1}(d–f)].

Specifically, we consider fluctuations in the threshold level,
$\xi^{(i)} \equiv \theta^{(i)}-\avg{\theta}$, as a primary source of
extrinsic noise. Such fluctuations may reflect variability in the
molecular components setting the division trigger or, more generally,
an effective activated process with a fluctuating barrier
height~\cite{Van_Kampen2007, Gardiner2009}.  We model the discrete-time dynamics of threshold
fluctuations as \begin{equation} \xi^{(i)} = \rho \xi^{(i-1)} +
  \eta^{(i)}, \ \ \text{with} \ \ \eta^{(i)}
  \overset{\text{i.i.d.}}{\sim} \mathcal{N}(0,\sigma_{\text{extr}}^2)
\label{eq:thredynamics}
\end{equation} where $\sigma^2_{\text{extr.}}$ represents extrinsic
noise and $\rho$ quantifies memory between successive cycles
[Fig.~\ref{fig:1}(b)–(c)]. Threshold (and size) fluctuations are
stable for $|\rho|<1$, with
$\var{\theta}=\sigma_{\text{extr}}^2/(1-\rho^2)$ and
$\covar{\theta^{(i)},\theta^{(i-k)}}=\rho^k\var{\theta}$. In what
follows, we restrict to $\rho \in (0, 1)$ for the half-reset case
($r=\tfrac{1}{2}$), and $\rho \in (-\tfrac{1}{2}, 1)$ for the
full-reset case ($r=0$), yielding the standard monotonic size-control
regime with $ \lambda_G \in (0, 1)$ (see below).  
For $\rho>0$, Eq.~(\ref{eq:thredynamics}) reduces to a discretely
sampled Ornstein–Uhlenbeck process with correlation time
$\tau_\xi = -\tau / \log \rho$, recovering previously studied
models~\cite{Luo2023}.

\begin{figure}[htpb]  \includegraphics[width=0.45\textwidth]{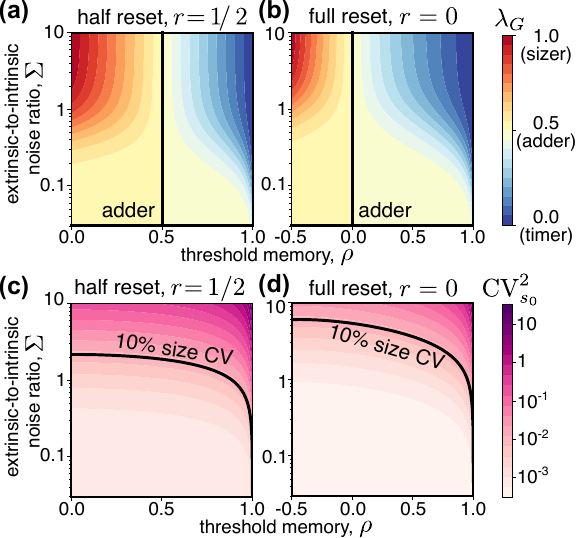}%
  \caption{\label{fig:2} \textbf{Model predictions.} Contour plots of
    the division-control parameter $\lambda_G$ (a,b) and birth-size
    fluctuations $CV^2(s_0)$ (c,d) from Eq.~(\ref{eq:fullsol}) with a plausible value of $\theta=1000$~\cite{Pla1991, Rueda2003, Ursinus2004, Mohammadi2009, Erickson2010, Li2014, Liu2015, Vischer2015, Egan2015, Ramey1978, Van_Heijenoort1992, Van_Heijenoort2007}, for $r=1/2$ (a,c) and $r=0$ (b,d).}
\end{figure}

Crucially, threshold fluctuations make the first-passage time in cycle
$(i)$ depend on the molecular increment required for division,
$\Delta x^{(i)}$, whose fluctuations obey
\begin{equation}
\Delta x^{(i)} - (1-r)\avg{\theta} = (\rho - r)\xi^{(i-1)} + \eta^{(i)}.
\label{eq:addedmolamlunt}
\end{equation}
To compute the division-control parameter and size fluctuations in
this regime, we can therefore use the expressions
$\lambda_G = - \alpha \avg{s_0} \tfrac{d \avg{\tau^{(i)} |
    s_0^{(i)}}}{ds_0^{(i)}}\Big|_{\avg{s_0}} = 1-\tfrac{1}{2}
\tfrac{\text{Cov}[s_f, s_0]}{\text{Var}[s_0]}$, and
$ \text{CV}^2_{s_0} = \frac{\alpha^2
  \avg{\var{\tau^{(i)}|s_0^{(i)}}}}{\lambda_G(2-\lambda_G)}, $ upon
applying the laws of total expectation and variance to
$\avg{\tau^{(i)}|s_0^{(i)},\Delta x^{(i)}}$ and
$\mathrm{Var}[\tau^{(i)}|s_0^{(i)},\Delta x^{(i)}]$. Remarkably, both
quantities can be written as their intrinsic limits
(Eq.~(\ref{eq:intrinsic-limit})), multiplied by a modulation depending
on two dimensionless parameters: (i) the threshold autocorrelation
$\rho$, and (ii) the relative strength of extrinsic to molecular
noise, $\Sigma \equiv \cv{\text{extr.}}/\cv{\text{mol.}}$, where
$\cv{\text{extr.}}^2=\sigma_{\text{extr}}^2/\avg{\theta}^2$ and
$\cv{\text{mol.}}^2=[(1-r)\avg{\theta}]^{-1}$.  The resulting
expressions take the compact form (see SI for analytical derivations)
\begin{equation} 
\begin{aligned}
    \frac{\lambda_G(\Sigma, \rho)}{\lambda_G^{\text{intr.}}} &= 1-\frac{1}{2} \frac{(\rho -r)(1-r\rho)}{(1-r)^2(1-\rho/2)} \frac{\text{CV}^2_{\theta}}{\text{CV}^2_{s_0}}(\Sigma, \rho),\\
     \frac{\text{CV}^2_{s_0}(\Sigma, \rho)}{(\text{CV}^2_{s_0})^\text{intr.}} &= 1+ \frac{(1+r^2)(1+\tfrac{\rho}{2})-r(1+2\rho)}{(1-r)^2(1-\rho/2)(1-\rho^2)} 
 \Sigma^2 ,
\end{aligned} \label{eq:fullsol}
\end{equation}with $
\text{CV}^2_{\theta}(\Sigma, \rho) = \frac{3 (\text{CV}^2_{s_0})^\text{intr.} \Sigma^2}{1-\rho^2}$. 
The full dependence of the division strategy ($\lambda_G$) and
birth-size fluctuations ($\mathrm{CV}^2_{s_0}$) on threshold memory
$\rho$, noise strength $\Sigma$, and reset fraction $r$ is summarized
in Fig.~\ref{fig:2}. Contrary to the common view that such models
generically implement a simple \emph{adder}, we find that they realize a
broad and continuously tunable range of behaviors. In particular, the
model captures the full spectrum of division strategies, from \emph{sizer} to
\emph{timer}, while allowing size fluctuations to vary from negligible to
experimentally relevant levels.

As expected, in the intrinsic-noise limit ($\Sigma \to 0$), the model
recovers \emph{adder} behavior~\cite{ElGamel2024}, which, however, is
associated with unrealistically small size fluctuations at typical
protein copy numbers. In addition to this known limit, we uncover a
qualitatively distinct \emph{adder} regime in which division control remains
\emph{adder} while size fluctuations are independently tunable. This occurs
when threshold memory matches the reset fraction ($\rho = r$),
defining a ``memory-balance'' [Fig.~\ref{fig:2}(a–b)], where
$\lambda_G = 1/2$ and
$\mathrm{CV}^2_{s_0}\to
\frac{1}{3(1-r)\langle\theta\rangle}\left(1+\frac{\Sigma^2}{1-r}\right)$. In
this regime, size variability increases continuously with threshold noise
without affecting division control [Fig.~\ref{fig:2}(c–d)], thereby
decoupling size fluctuations from the underlying control strategy.
This decoupling originates from a cancellation of memory effects
(Eq.~(\ref{eq:addedmolamlunt})): when $\rho=r$, correlations in the
threshold exactly compensate the memory introduced by protein
inheritance at division, rendering the accumulated increment
statistically independent across cycles and thereby restoring \emph{adder}
control. Thus, \emph{adder} behavior does not require full protein
reset. Instead, full reset ($r=0$) yields \emph{adder} control only when
threshold fluctuations are uncorrelated ($\rho=0$), whereas in general
the condition is $\rho=r$. This revises the standard intuition that
full protein reset is necessary for \emph{adder} control in presence of
threshold noise.
We note that this behavior contrasts with models such as
Ref.~\cite{ElGamel2024}, in which increasing size fluctuations via
degradation or constant production terms simultaneously alters the
division strategy, and that in those models degradation can be seen as
a contributor to memory-balance.

Away from the $\rho=r$ line, increasing threshold noise drives the
system away from the intrinsic \emph{adder} limit ($\Sigma\to0$), shifting
division control toward either sizer-like ($\rho<r$) or timer-like
behavior ($\rho>r$) [Fig.~\ref{fig:2}(a)]. Access to both regimes
requires partial inheritance of the divisor molecule ($r>0$). In the
fully reset case ($r=0$), division control is instead confined to
interpolate between \emph{adder} and \emph{timer}, except when negatively correlated
fluctuations are allowed ($\rho<0$), which enable recovery of
sizer-like behavior [Fig.~\ref{fig:2}(b)]. Specifically, in the large-noise limit ($\Sigma\gg1$), the asymptotic
behavior reads
$\frac{\text{CV}^2_{s_0}}{\text{CV}^2_{\theta}} \to
\frac{(1+r^2)(1+\rho/2)-r(1+2\rho)}{3(1-r)^2 (1-\rho/2)}$ and
$\lambda_G \to \tfrac{1}{2} \lbrace 1+ \tfrac{3}{2}
\frac{(\rho-r)(\rho r-1)}{(1+r^2)(1+\rho/2)-r(1+2\rho)} \rbrace$. For
$r=\tfrac{1}{2}$, one obtains
$\text{CV}^2_{s_0} \to \text{CV}^2_{\theta}$ independently of $\rho$
and $\lambda_G \to 1-\rho$, so that division control spans the full
range from perfect \emph{sizer} ($\rho=0$) to perfect \emph{timer} ($\rho=1$). By
comparison, for $r=0$,
$\text{CV}^2_{s_0} \to \frac{2+\rho}{3 (2-\rho)}\text{CV}^2_{\theta}$
and $\lambda_G \to \tfrac{1}{2}\left(1-\tfrac{3\rho}{2+\rho}\right)$,
yielding a continuous interpolation from perfect \emph{sizer} at $\rho=-1/2$
to perfect \emph{timer} at $\rho=1$.

In the fully autocorrelated limit ($\rho \to 1$), the model approaches
a perfect \emph{timer}, with $\mathrm{CV}^2_{s_0} \to \mathrm{CV}^2_\theta$
and $\lambda_G \to 0$, independently of $\Sigma$ and $r$, provided
$\Sigma \neq 0$. This may appear paradoxical, since in this limit
$\theta^{(i)} \simeq \theta^{(i-1)} + \eta^{(i)}$, implying that over
finite timescales its deviations from the initial value remain small
with high probability, especially when extrinsic noise is weak.  More
precisely, for any finite extrinsic noise, there exists a
characteristic number of generations $\mathcal{L}$ over which the
threshold remains within a fraction $\epsilon$ of its initial value
$\theta^0$ with confidence $p$:
$\mathcal{L} \approx \left(\frac{\epsilon
    \theta^0}{z_{(1+p)/2}\sigma_{\text{extr.}}}\right)^2$. Over such
timescales, the threshold is effectively constant, which would suggest
adder-like behavior. The resolution lies in variability across
lineages. Individual lineages behave approximately as \emph{adder}s for
$\lesssim \mathcal{L}$ generations, but differences in the inherited
threshold $\theta^0$ dominate at longer times and give rise to an
emergent \emph{timer} at the population level (SI Fig.~2). This instance
of Simpson's paradox highlights a nontrivial interplay between lineage
and population behavior.

\begin{figure}[htpb]  \includegraphics[width=0.47\textwidth]{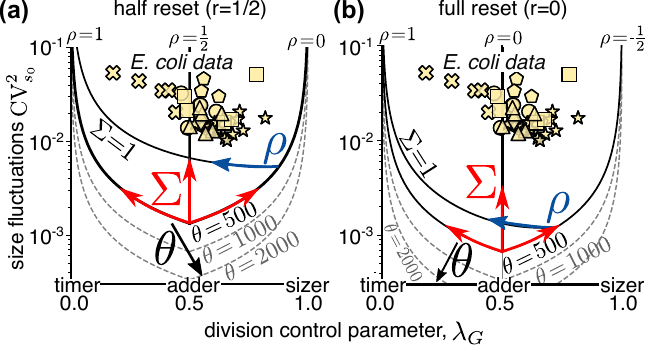}%
  \caption{\label{fig:3} \textbf{Extrinsic sources of noise are needed
      to fit \ecoli data at realistic divisor molecule numbers.} (a–b)
    Predicted joint modulation of birth-size fluctuations
    ($\mathrm{CV}^2(s_0)$) and division control ($\lambda_G$) as model
    parameters ($\Sigma, \rho, \theta$) vary, for partial reset
    ($r=1/2$, a) and full reset ($r=0$, b). (c) Single-cell
    measurements from seven independent microfluidic experiments on
    \ecoli cluster near the \emph{adder} regime ($\lambda_G \simeq 0.5$) with
    birth-size fluctuations of $\sim10$–$20\%$.  }
\end{figure}

Finally, to compare the model predictions with available \textit{E.~coli}
single-cell data, we focused on the predicted relationship between size fluctuations,
$\mathrm{CV}^2(s_0)$, and the division-control parameter, $\lambda_G$, and sourced measurements from seven independent
studies~\cite{Taheri-Araghi2015, Adiciptaningrum2015, Kennard2016,
  Wallden2016, Si2019, Witz2019, Colin2021}.
Consistent with previous analyses, the data cluster near the \emph{adder}
regime ($\lambda_G \simeq 0.5$), with a weak shift toward sizer-like
behavior at low growth rates, and exhibit birth-size fluctuations in
the range $\mathrm{CV}^2(s_0) \sim 10^{-2} - 4 \times 10^{-2}$ [Fig.~\ref{fig:3}].  Comparing these trends with the full range of model
predictions obtained by varying $(\rho, \Sigma, r, \theta)$
[Fig.~3(a),(b)], we first observe that decreasing the threshold level
shifts the entire manifold upward by increasing the intrinsic baseline
of size fluctuations. At biologically realistic molecule numbers
($\sim 10^3$~\cite{Pla1991, Rueda2003, Ursinus2004, Mohammadi2009, Erickson2010, Li2014, Liu2015, Vischer2015, Egan2015, Ramey1978, Van_Heijenoort1992, Van_Heijenoort2007}), the model nevertheless cannot access the experimental
region without extrinsic noise, indicating that extrinsic noise is
essential to reproduce the observed size statistics.

In the presence of extrinsic noise,  agreement with
the data is achieved in a memory-balanced \emph{adder} regime,
$\rho \approx r$. In this regime, threshold fluctuations modulate size
variability while maintaining division control close to the observed
adder. When the memory-balanced condition is perturbed, $\rho > r$ promotes timer-like behavior and $\rho < r$ promotes
sizer-like behavior. (The well-studied scenario of full resetting is a particular case of this class of models, for which $\rho=r=0$.)
Additional biological ingredients not included in our main framework,
such as active protein degradation or constant production, also tune
division control and size fluctuations (SI Fig.~1), extending the
accessible regimes~\cite{ElGamel2024}.


In this work, we investigated the role of extrinsic noise in a
stochastic threshold–accumulation model for cell division. Contrary to
common assumptions~\cite{ElGamel2024, Panlilio2021,Si2019}, we uncover
a phenomenology that extends beyond simple \emph{adder} behavior. By solving
the model analytically, we derived explicit relationships linking
division control to size fluctuations, connecting underlying
stochastic molecular processes to emergent phenomenological laws.
We should note that although stochastic accumulation rules with threshold variability and memory already capture a rich phenomenology, chromosome replication and segregation are also known to play central roles in \emph{E.~coli} cell-division control~\cite{Micali2018a,Micali2018b,Colin2021,Tiruvadi-Krishnan2022,Ho2015}.

While incorporating threshold-accumulation processes into a more complete theory remains an important challenge, our analysis leads to three key results. First, intrinsic stochasticity
alone, although it naturally generates \emph{adder}-like correlations,
predicts size fluctuations that are significantly smaller than those
observed in \textit{E.~coli} for realistic threshold molecule
numbers~\cite{Pla1991, Rueda2003, Ursinus2004, Mohammadi2009, Erickson2010, Li2014, Liu2015, Vischer2015, Egan2015, Ramey1978, Van_Heijenoort1992, Van_Heijenoort2007}. A systematic exploration of the model shows that extrinsic
variability—specifically, fluctuations in the division threshold—is an
essential ingredient for quantitatively capturing experimental
observations within a threshold-accumulation framework.

Second, we connect the size-control parameter to the
  amplitude and memory of the extrinsic noise. \emph{Adder} behavior emerges
  when threshold memory compensates for molecular reset at division,
  i.e., the retained fraction of the divisor protein, independent of
  noise amplitude.

Finally, our results show that the \emph{adder} principle does not, in general, minimize size variability in the presence of extrinsic noise, as the optimal control strategy becomes explicitly noise-dependent. In this setting, constrained optimization does not provide a well-defined route to an optimality principle, since relative fluctuations can always be reduced by tuning the threshold. Even in the absence of extrinsic noise, optimization over a restricted set of parameters (e.g., the degradation rate alone) lacks a clear biological basis, as evolutionary processes can act on multiple control variables. Taken together, these findings challenge the view that \emph{adder} correlations arise from variability minimization and instead point to a broader, noise-dependent landscape of size-control strategies. Whether minimizing size variability is itself under selection remains unclear~\cite{Lin2017StochasticityPopulationGrowth, Levien2021NonGeneticVariability, Hobson-Gutierrez2025, Proulx-Giraldeau2022}, leaving the origin of the widely observed \emph{adder} correlations an open question.

We thank Gabriele Micali for useful discussions.
This work was supported
by Fondazione AIRC per la ricerca sul cancro ETS (AIRC IG 2019 Grant
No. 23258 and AIRC IG 2024 Grant No. 30391—P.I. M. C.
L.). M.C. was supported by Fondazione AIRC per la ricerca sul
cancro ETS (ID. 28177). A.A. and K.B. were supported by the European Union (ERC, BIGR, 101125981) and Israeli Science Foundation (146873).

\bibliography{references}

\end{document}


\begin{center}
{\LARGE
Supplementary Information for\\ 
``Essential Role of Extrinsic Noise in Models of \emph{E.~coli} Division Control''
}
\end{center}

\vspace{1em}

\section{Phenomenological framework for cell division control}
We begin by briefly reviewing key concepts and quantities in the phenomenological study of division
control~\cite{Amir2014, Ho2018, Grilli2018}. 

At the phenomenological level, division control in
exponentially growing cells is described by a control function
$\delta G^{(i)}(\delta q_0^{(i)})$, linking fluctuations in the added
logarithmic size $G^{(i)} \equiv \log (s_f^{(i)}/s_0^{(i)})$ to those
in the logarithmic birth size $q_0^{(i)} \equiv \log (s_0^{(i)}/s^*)$ at cell cycle $(i)$~\cite{Amir2014}.  

Motivated by the smallness of fluctuations around the mean, the
control function is commonly linearized~\cite{Amir2014,Ho2018,Grilli2018} as $\delta G^{(i)} \approx -\lambda_G \, \delta q_0^{(i)} +
\eta_G^{(i)}$, where \begin{equation}
    \lambda_G  \equiv - \frac{d\,\avg{G^{(i)} | q_0^{(i)}}}{d q_0^{(i)}}\Big|_{q_0^{(i)} = \mathbb{E}[q_0]} \label{eq:lambdaG}.
\end{equation} Importantly, the division control parameter $\lambda_G$ introduced by this linear approximation can be directly estimated from the data
as \begin{equation}
    \lambda_G  = - \frac{\covar{G^{(i)}, q_0^{(i)}}}{ \var{q_0}},
\end{equation}
and used to distinguish between different division control strategies. Alternatively, one may consider the control function
$\delta \Delta s^{(i)}(\delta s_0^{(i)})$, defined in terms of the
added size $\Delta s^{(i)} \equiv s_f^{(i)} - s_0^{(i)}$, with the
corresponding control parameter \begin{equation}
  \zeta_{\Delta} \equiv - \frac{d\,\avg{\Delta s^{(i)} | s_0^{(i)}}}{d
  s_0^{(i)}}\Big|_{s_0^{(i)} = \mathbb{E}[s_0]}= - \frac{\covar{\Delta
    s^{(i)}, s_0^{(i)}}}{ \var{s_0}}.  
\end{equation} The two formulations are equivalent to first order, describing the
same underlying control mechanism in different variables, and are
related by~\cite{Micali2018b} \begin{equation}
    \lambda_G = 1 - \frac{\avg{s_0}}{\avg{s_f}} (1 - \zeta_\Delta) = 1 - \frac{1}{2} (1-\zeta_\Delta), \label{eq:equiv}
\end{equation}where the last equality holds for symmetric division. For an
adder, cells add on
average a constant, size-independent increment
$s_f^{(i)} = s_0^{(i)}+\Delta+\text{noise}$, hence $\zeta_\Delta = 0$ and
$\lambda_G = 1-\avg{s_0} / \avg{s_f}$. Furthermore, if division is symmetric, $\lambda_G = \tfrac{1}{2}$ for an adder.

Within this framework, the logarithmic birth size follows a
discrete-time first order auto-regressive process \begin{equation}
    \delta q_0^{(i)} = (1-\lambda_G)\delta q_0^{(i-1)}+\eta^{(i-1)},
\end{equation} converging to a stationary distribution for $0<\lambda_G<2$
(equivalently,
$1-\frac{\avg{s_f}}{\avg{s_0}}<\zeta_\Delta<1+\frac{\avg{s_f}}{\avg{s_0}}$)~\cite{Amir2014,Ho2018}.
In this limit, birth-size fluctuations satisfy
\begin{equation}
    \cv{s_0} = \sqrt{\frac{\var{\eta_G}}{\lambda_G (2-\lambda_G)}} 
    = \sqrt{\frac{\avg{\var{G^{(i)} | q_0^{(i)
    }}}}{\lambda_G (2-\lambda_G)}},  \label{eq:cvs0}
\end{equation}
and are (apparently!~\cite{ElGamel2024}) minimized by a ``sizer'' ($\lambda_G=\zeta_\Delta = 1$).

\section{Division control in presence of molecular noise}
Considering molecular noise on the accumulation of the divisor molecule ``x'', synthesized at a rate $\mu s_0 e^{\alpha t}$, division time becomes the first passage time of a non-homogenous Poisson process (NHPP), $x(t)$, to a fixed threshold $\theta$, starting from an initial amount $r \theta$: \begin{equation}
    \tau^{(i)} = \inf \lbrace  t> 0 : x^{(i)}(t) - x_0 \geq (1-r) \theta \rbrace.
\end{equation} 

The first passage time statistics are related to the division control parameter, $\lambda_G$, and the birth size fluctuations, $\text{CV}^2_{s_0}$, through Eq.~(\ref{eq:lambdaG}) and Eq.~(\ref{eq:cvs0}), respectively, and the relation $G^{(i)}=\alpha^{(i)} \tau^{(i)} $, valid under exponential growth. Assuming a constant growth rate, $\alpha$, we have \begin{equation}
     \lambda_G = -
  \alpha \avg{s_0} \frac{d \avg{\tau^{(i)} |
      s_0^{(i)}}}{ds_0^{(i)}}\Big|_{\avg{s_0}}, \quad \text{CV}^2_{s_0} =
  \frac{\alpha^2 \avg{\var{\tau^{(i)}|s_0^{(i)}}}}{\lambda_G(2-\lambda_G)}. \label{eq:connection}
\end{equation} 

The conditional expectation and variance of the FPT of a NHPP can be written as~\cite{Van_Kampen2007, Gardiner2009}\begin{equation}
    \mathbb{E}[\tau^{(i)} | s_0^{(i)}] = \frac{1}{\alpha} \mathbb{E}\left[ \log \left( 1+ \frac{\alpha}{\mu s_0^{(i)}} Y \right) \right]; \quad \text{Var}[\tau^{(i)} | s_0^{(i)}] = \frac{1}{\alpha^2} \text{Var}\left[ \log \left( 1+ \frac{\alpha}{\mu s_0^{(i)}} Y \right) \right],
\end{equation}  $\text{with} \ \ Y\sim \text{Gamma}((1-r)\theta, 1)$. For large threshold values, $\theta\gg 1$, we can exploit the normal approximation, $Y \approx  (1-r)\theta + \sqrt{(1-r) \theta } \  Z$, with $Z\sim \mathcal{N}(0, 1)$, to write \begin{equation}
    \log \left(1+ \frac{\alpha}{\mu s_0^{(i)}} Y \right) \approx \log \left(1+ \frac{\alpha}{\mu s_0^{(i)}} (1-r)\theta\right) + \frac{\frac{\alpha}{\mu s_0^{(i)}}\sqrt{(1-r)\theta}}{1+ \frac{\alpha}{\mu s_0^{(i)}} (1-r)\theta} Z.
\end{equation} Therefore,
\begin{equation}
    \mathbb{E}[\tau^{(i)} | s_0^{(i)}] = \frac{1}{\alpha}\log \left(1+ \frac{\alpha}{\mu s_0^{(i)}} (1-r)\theta\right); \quad \text{Var}[\tau^{(i)} | s_0^{(i)}] = \frac{1}{\alpha^2}  \frac{\left(\frac{\alpha}{\mu s_0^{(i)}}\right)^2 (1-r)\theta}{\left(1+ \frac{\alpha}{\mu s_0^{(i)}} (1-r)\theta\right)^2}.
\end{equation}

Using Eq.~(\ref{eq:connection}), we have \begin{equation}
    \lambda_G = -\alpha \mathbb{E}[s_0] \frac{1}{\alpha} \frac{-\frac{\alpha}{\mu \mathbb{E}[s_0]^2} (1-r)\theta}{1+\frac{\alpha}{\mu \mathbb{E}[s_0]} (1-r)\theta} = \frac{\frac{\alpha}{\mu} (1-r)\theta}{\mathbb{E}[s_0] + \frac{\alpha}{\mu} (1-r)\theta} = \frac{\mathbb{E}[\Delta s]}{\mathbb{E}[s_f]} = \frac{1}{2},
\end{equation} and \begin{equation}
    \text{CV}^2_{s_0} = \frac{\alpha^2}{3} \frac{1}{\alpha^2}  \frac{\left(\frac{\alpha}{\mu \mathbb{E}[s_0]}\right)^2 (1-r)\theta}{\left(1+ \frac{\alpha}{\mu \mathbb{E}[s_0]} (1-r)\theta\right)^2} \approx \frac{1}{3 (1-r) \theta }.
\end{equation} These corresponds to the intrinsic limit solutions considered in the main text.

An alternative and equivalent formula for the division control parameter can be derived noting that the conditional expected value of the first-passage time satisfies \begin{equation}
    \avg{\tau^{(i)} |
      s_0^{(i)}} = \avg{\tau} + \frac{\covar{\tau^{(i)}, s_0^{(i)}}}{\var{s_0}} (s_0^{(i)} - \avg{s_0}). \label{eq:connection2}
\end{equation}
Considering exponential growth, $s_f^{(i)} = s_0^{(i)} e^{\alpha \tau^{(i)}}$, and keeping only linear-order fluctuations, one has: \begin{equation}
    \avg{\tau^{(i)} | s_0^{(i)}} = \frac{1}{\alpha} \log \left( \frac{\avg{s_f^{(i)} | s_0^{(i)}}}{s_0^{(i)}} \right),
\end{equation}  and \begin{equation}
\covar{\tau^{(i)},s_0^{(i)}} = \frac{1}{\alpha \avg{s_0}} \left( \frac{\covar{s_f^{(i)},s_0^{(i)}} }{e^{\alpha \avg{\tau}}} - \var{s_0}\right) .
\end{equation} Inserting this expression into Eq.~(\ref{eq:connection2}) and using Eq.~(\ref{eq:connection}) we find \begin{equation}
    \lambda_G = 1- \frac{1}{2}\frac{\covar{
    s_f^{(i)}, s_0^{(i)}}}{ \var{s_0}}.
\end{equation}

\section{Effects of threshold fluctuations on division control}\label{sec:div_controll_cor_th}
As discussed in the main text, in the presence of threshold fluctuations \begin{equation}\label{eq:Th_dy_AR1}
\theta^{i} = (1-\rho) \langle \theta \rangle + \rho \theta^{i-1} + \eta^{i},
\qquad  
\eta^{i} \stackrel{\text{i.i.d.}}{\sim}\mathcal{N}(0,\sigma_{\text{extr.}}^{2}),
\end{equation} with \begin{equation}\label{eq:threshold_AR1}
\mathrm{Var}(\theta^{i})=\sigma_{\theta}^{2} = \sigma_{\text{extr.}}^{2}/(1-\rho^2),\qquad
\mathrm{Cov}(\theta^{i-k},\theta^{i})=\rho^{k}\sigma_{\theta}^{2},
\end{equation} Eq.~(1) of the main text should be replaced with 
\begin{equation}\label{eq:s_f}
    s_f^{i} = s_0^{i} + \frac{\alpha}{\mu} (\theta^{i}-r\theta^{i-1}), \qquad s_0^i = \frac{s_f^{i-1}}{2}.
\end{equation} We will now use Eq.~(\ref{eq:s_f}) to derive the expressions for size fluctuations and division control parameter reported in Eq.~(5) of the main text.

\subsection*{Linear regression between birth and division size}
We start by computing the slope of the linear regression between birth and division size, defined as $1-\zeta_\Delta \;=\; \mathrm{Cov}(s_f^i,s_0^i) / \sigma^2_{s_0}$.

Using Eq.~(\ref{eq:s_f}),
\begin{equation}
    \mathrm{Cov}(s_f^i,s_0^i)
    = \sigma^2_{s_0} + \frac{\alpha}{\mu}\left[\mathrm{Cov}(\theta^{i},s_0^i)-r\,\mathrm{Cov}(\theta^{i-1},s_0^i)\right],
    \label{eq:cov_sf_s0_expand}
\end{equation} so that it remains to evaluate $\mathrm{Cov}(\theta^{i},s_0^i)$ and $\mathrm{Cov}(\theta^{i-1},s_0^i)$.
\paragraph*{Calculation of $\mathrm{Cov}(\theta^{i},s_0^i)$.} To compute this term, we use symmetric division, $s_0^i=s_f^{i-1}/2$, and Eq.~(\ref{eq:s_f}), to write
\begin{align}
    \mathrm{Cov}(\theta^{i},s_0^i)
    &= \frac{1}{2}\,\mathrm{Cov}\!\left(\theta^{i},\,s_f^{i-1}\right)\nonumber
    = \frac{1}{2}\,\mathrm{Cov}\!\left(\theta^{i},\,s_0^{i-1} + \frac{\alpha}{\mu}\bigl(\theta^{i-1}-r\theta^{i-2}\bigr)\right)\nonumber\\
    &= \frac{1}{2}\,\mathrm{Cov}(\theta^{i},s_0^{i-1})
    +\frac{1}{2}\frac{\alpha}{\mu}\left[\mathrm{Cov}(\theta^{i},\theta^{i-1})
    -r\,\mathrm{Cov}(\theta^{i},\theta^{i-2})\right].
    \label{eq:cov_X_s0_step1}
\end{align}
Now, using the threshold statistics of Eq.~(\ref{eq:threshold_AR1}), we obtain
\begin{equation}
    \mathrm{Cov}(\theta^{i},s_0^i)
    \;=\; \frac{1}{2}\,\mathrm{Cov}(\theta^{i},s_0^{i-1})
    +\frac{1}{2}\frac{\alpha}{\mu}\left(\rho-r\rho^2\right)\sigma_{\theta}^2.
    \label{eq:cov_X_s0_step2}
\end{equation}
Finally, using the discrete-time AR(1) threshold dynamics in Eq.~(\ref{eq:Th_dy_AR1}) and the fact that $\eta^i$ is independent of $s_0^{i-1}$, we can write
\begin{equation}
    \mathrm{Cov}(\theta^{i},s_0^{i-1})
    = \rho\,\mathrm{Cov}(\theta^{i-1},s_0^{i-1}),
\end{equation}
so that Eq.~(\ref{eq:cov_X_s0_step2}) becomes the recursion
\begin{equation}
    \mathrm{Cov}(\theta^{i},s_0^i)
    \;=\; \frac{1}{2}\rho\,\mathrm{Cov}(\theta^{i-1},s_0^{i-1})
    +\frac{1}{2}\frac{\alpha}{\mu}\left(\rho-r\rho^2\right)\sigma_{\theta}^2.
    \label{eq:cov_X_s0_recursion}
\end{equation}
In the stationary regime, $\mathrm{Cov}(\theta^{i},s_0^i)=\mathrm{Cov}(\theta^{i-1},s_0^{i-1})$, and solving Eq.~(\ref{eq:cov_X_s0_recursion}) yields
\begin{equation}\label{eq:cov_1}
    \mathrm{Cov}(\theta^{i},s_0^i)
    \;=\;
    \frac{1}{2}\frac{\alpha}{\mu}\,
    \frac{\rho-r\rho^2}{1-\rho/2}\,
    \sigma_{\theta}^2.
\end{equation}
\paragraph*{Calculation of $\mathrm{Cov}(\theta^{i-1},s_0^i)$:}
Following similar algebraic steps, and using the AR(1) relation $\mathrm{Cov}(\theta^{i-1},s_0^{i-1})=\mathrm{Cov}(\theta^{i},s_0^{i})$ in stationarity, we obtain
\begin{eqnarray}\label{eq:cov_2}
    \mathrm{Cov}(\theta^{i-1},s_0^i)
    =
    \frac{1}{2}\,\mathrm{Cov}(\theta^{i-1},s_0^{i-1})
    + \frac{1}{2}\frac{\alpha}{\mu}\left(1-r\rho\right)\sigma_{\theta}^2
    =\frac{1}{2}\frac{\alpha}{\mu}\frac{1-r\rho}{1-\rho/2}\,\sigma_{\theta}^2.
\end{eqnarray}
Now substituting Eqs.~(\ref{eq:cov_1}) and (\ref{eq:cov_2}) into Eq.~(\ref{eq:cov_sf_s0_expand}), we obtain
\begin{equation}
    \mathrm{Cov}(s_f^i,s_0^i)
    \;=\;
    \sigma^2_{s_0}
    +\frac{1}{2}\Big(\frac{\alpha}{\mu}\Big)^2\sigma_{\theta}^2
    \frac{(1-r\rho)(\rho-r)}{1-\rho/2}.
\end{equation}
Hence, in a stationary condition, the slope of the linear regression between birth and division size becomes
\begin{equation}
    1-\zeta_\Delta \;=\; 1+\frac{1}{2}\Big(\frac{\alpha}{\mu}\Big)^2 \frac{\sigma_{\theta}^2}{\sigma_{s_0}^2}
    \frac{(1-r\rho)(\rho-r)}{1-\rho/2}.
    \label{eq:zetaG_general}
\end{equation}

\subsection*{Size fluctuations}
As explained in the text, to compute $\sigma^2_{s_0}$ in the presence of threshold fluctuations, we can use the law of total variance
\begin{equation}\label{eq:sig_total}
\sigma^2_{s_0}
\;=\;
\big\langle \sigma^2_{s_0|\theta}\big\rangle
+\sigma^2_{\langle s_0|\theta\rangle},
\end{equation}
where $\sigma^2_{s_0|\theta}$ denotes the variance of $s_0$ conditioned on a fixed threshold, and $\sigma^2_{\langle s_0|\theta\rangle}$ is the variance of the conditional mean across threshold realizations.

\paragraph*{Calculation of first term.} The birth-size variance at a fixed threshold, coming from intrinsic molecular fluctuations, has already been calculated in \cite{ElGamel2024}, yielding
\begin{equation} \label{eq:sig_int}
\big\langle \sigma^2_{s_0|\theta}\big\rangle
\;=\;
\frac{\alpha}{\mu}\,\frac{\langle s_0\rangle}{3} = 
\Big(\frac{\alpha}{\mu}\Big)^2 \frac{(1-r)\,\langle \theta\rangle}{3}, 
\end{equation}where for the last equality, we have substituted the expression $ \langle s_0\rangle = \frac{\alpha}{\mu}(1-r)\langle \theta\rangle.$

\paragraph*{Calculation of second term.}
Next, we compute $\sigma^2_{\langle s_0 \mid \theta\rangle}
\equiv \mathrm{Var}\Big(\,\langle s_0^i \mid \{\theta\}\rangle\Big)$. Using Eq.~(\ref{eq:s_f}) and setting $m_i \equiv \langle s_0^i \mid \{\theta\}\rangle$, we obtain
\begin{align}
m_i&=\frac12 m_{i-1}+\frac{\alpha}{2\mu}\Big(\theta^{i-1}-r\theta^{i-2}\Big).
\label{eq:mi_recursion}
\end{align}

Let $\sigma_m^2\equiv \mathrm{Var}(m_i)$ in stationarity. Taking the variance of
Eq.~\eqref{eq:mi_recursion} gives
\begin{align}
\sigma_m^2
&=\frac{1}{3}\left(\frac{\alpha}{\mu}\right)^2 \mathrm{Var}\!\Big(\theta^{i-1} -r\theta^{i-2} \Big)
+\frac{2}{3}\frac{\alpha}{\mu}\Big(\mathrm{Cov}(m_{i-1},\theta^{i-1})-r\,\mathrm{Cov}(m_{i-1},\theta^{i-2})\Big).
\label{eq:sigmam_rearranged}
\end{align}
Substituting Eqs.~\eqref{eq:cov_1} and \eqref{eq:cov_2} into
Eq.~\eqref{eq:sigmam_rearranged} and solving for $\sigma_m^2$ yields
\begin{align}
\sigma^2_{\langle s_0 \mid \theta\rangle}
&=\frac{1}{3}\left(\frac{\alpha}{\mu}\right)^2\sigma_{\theta}^2
\left[
(1+r^2-2r\rho)+\frac{(\rho-r)(1-r\rho)}{(1-\rho/2)}
\right].
\label{eq:sig_ext}
\end{align}
Putting all together, we obtain the total birth size variance as
\begin{equation}\label{eq:var_total}
    \sigma_{s_0}^2 = \frac{1}{3}\left(\frac{\alpha}{\mu}\right)^2\left((1-r)\,\langle \theta\rangle+\sigma_{\theta}^2\left[(1+r^2-2r\rho)+\frac{(\rho-r)(1-r\rho)}{(1-\rho/2)}\right]\right).
\end{equation}

\paragraph*{Variance in terms of relative threshold to molecular noise.} Following the main text, we introduce a dimensionless noise ratio that quantifies the relative strength of extrinsic noise from threshold fluctuations to intrinsic noise from molecular fluctuations, given as follows
\begin{equation}
    \Sigma^2 \equiv \frac{\mathrm{CV}^2_{\text{extr.}}}{\mathrm{CV}^2_{\text{mol}}} = (1-\rho^2)\frac{\mathrm{CV}^2_{\theta}}{\mathrm{CV}^2_{\text{mol}}} = (1-\rho^2)\tilde{\Sigma}^2,
\end{equation}
where $\mathrm{CV}^2_{\theta} = \sigma^2_{\theta}/\langle \theta\rangle^2$ and
$\mathrm{CV}^2_{\text{mol}} = \big\langle \sigma^2_{s_0|\theta}\big\rangle/\langle s_0\rangle^2 = 1/[(1-r)\langle \theta\rangle]$. In terms of this dimensionless quantity the birth size variance can be written as:
\begin{equation}
    \sigma_{s_0}^2 = \frac{\sigma_{\theta}^2}{3}\left(\frac{\alpha}{\mu}\right)^2\left(\frac{(1-r)^2}{\tilde{\Sigma}^2}+(1+r^2-2r\rho)+\frac{(\rho-r)(1-r\rho)}{(1-\rho/2)}\right). \label{eq:vars0}
\end{equation}
Dividing Eq.~(\ref{eq:vars0}) by $\langle s_0 \rangle^2$ we can get the expression for the coefficient of variation of birth size fluctuations as \begin{equation}
    \text{CV}^2_{s_0} = \text{CV}^2_{mol} \left[ 1  +  \frac{(1+r^2-2r\rho)(1-\rho/2)+(\rho-r)(1-r\rho)}{(1-r)^2(1-\rho/2)} \tilde{\Sigma}^2 \right].
\end{equation} One can easily check that this expression is equivalent to the one reported in the main text.
\subsection*{Division control parameter}
To obtain the expression for the division control parameter, we can substitute Eq.~(\ref{eq:vars0}) into Eq.~(\ref{eq:zetaG_general}) to write
\begin{equation}\label{eq:zeta_final}
1-\zeta_\Delta=1+\frac{3}{2}\,\frac{(\rho-r)(1-r\rho)}{(\rho-r)(1-r\rho)+\left(1-\rho/2\right)\left[\frac{(1-r)^2}{\tilde{\Sigma}^2}+(1+r^2-2r\rho)\right]},
\end{equation}
from which, using Eq.~(\ref{eq:equiv}), we have:
\begin{equation}\label{eq:Div_controll_final}
    \lambda_G = \frac{1}{2}\left(1-\frac{3}{2}\,\frac{(\rho-r)(1-r\rho)}{(\rho-r)(1-r\rho)+\left(1-\rho/2\right)\left[\frac{(1-r)^2}{\tilde{\Sigma}^2}+(1+r^2-2r\rho)\right]}\right),
\end{equation}which is equivalent to the one reported in the main text.

\section{Effects of protein reset fraction fluctuations}
For simplicity, in the following analysis we assume $\rho = 0$. By following the same steps leading to Eq.~(\ref{eq:zetaG_general}), one finds that, in the presence of noise in the protein reset fraction $r$, the expression for $1-\zeta_\Delta$ in the case of uncorrelated thresholds remains identical to Eq.~(\ref{eq:zetaG_general}) evaluated at $\rho = 0$:
\begin{equation}
1-\zeta_\Delta =
1-\frac{r}{2}
\left(
\frac{\alpha}{\mu}\cdot\frac{\sigma_\theta}{\sigma_{s_0}}
\right)^2 . \label{SIeq:slopewithr}
\end{equation}
However, the expression for the variance of the birth size differs in this case. For uncorrelated thresholds, using Eq.~(\ref{eq:s_f}) we obtain
\begin{eqnarray}
3\sigma^2_{\langle s_0 \mid \theta\rangle}
&=&
\left(\frac{\alpha}{\mu}\right)^2
\left[
\mathrm{Var}(\theta)
+
\mathrm{Var}(r\theta)
\right]
-
2\langle r\rangle
\left(\frac{\alpha}{\mu}\right)
\mathrm{Cov}(s_0^i,\theta^{i-1}),
\\
\sigma^2_{\langle s_0 \mid \theta\rangle}
&=&
\left(\frac{\alpha}{\mu}\right)^2
\frac{\sigma_\theta^2}{3}
\left[
1+\langle r\rangle^2-\langle r\rangle
+
\sigma_r^2
\left(
1+\frac{\langle\theta\rangle^2}{\sigma_\theta^2}
\right)
\right],
\end{eqnarray}
where Eq.~(\ref{eq:cov_2}) has been used for $\mathrm{Cov}(s_0^i,\theta^{i-1})$ in the limit $\rho\to0$. Applying the law of total variance and combining with Eq.~(\ref{eq:sig_int}), we obtain
\begin{equation}\label{eq:sig_b_noise}
\sigma_{s_0}^2
=
\left(\frac{\alpha}{\mu}\right)^2
\left[
(1-r)\frac{\langle\theta\rangle}{3}
+
\left(
1+\langle r\rangle^2-\langle r\rangle
+
\sigma_r^2
\left(
1+\frac{\langle\theta\rangle^2}{\sigma_\theta^2}
\right)
\right)
\frac{\sigma_\theta^2}{3}
\right].
\end{equation}
In the absence of noise in the protein reset fraction, i.e.\ for $\sigma_r=0$, the expression for $\sigma_{s_0}^2$ reduces to Eq.~(\ref{eq:var_total}) evaluated at $\rho=0$. In contrast, increasing noise in the protein reset fraction reduces the correction term in Eq.~(\ref{SIeq:slopewithr}). Therefore, while threshold noise drives the system away from the intrinsic adder behavior, fluctuations in the protein reset fraction act to restore a more adder-like behavior.

\section{Effects of constant degradation of divisor protein}
Including a degradation term $-\gamma x^{(i)}(t)$, in the synthesis dynamics of the divisor protein modifies Eq.~(\ref{eq:s_f}), introduced in Section~\ref{sec:div_controll_cor_th}, which now takes the more general form
\begin{eqnarray}\label{eq:eq:s_f_degradation}
s_f^i
&=&
s_0^i 2^{-\epsilon}
+
\frac{\alpha}{\mu}(1+\epsilon)
\left(
\theta^{i}
-
r\theta^{i-1}2^{-\epsilon}
\right),
\end{eqnarray}
where $\epsilon=\gamma/\alpha$ denotes the ratio between the degradation rate and the cellular growth rate. In the limit of $\epsilon \to 0$, the above Eq.~(\ref{eq:eq:s_f_degradation}) converges to Eq.~(\ref{eq:s_f}). We will now use Eq.~(\ref{eq:eq:s_f_degradation}) to derive the expressions for size fluctuations and division control parameter. For simplicity, we assume $\rho = 0$ in the subsequent calculations.

\subsection*{Linear regression between birth and division size}
Following the same steps as in Section~\ref{sec:div_controll_cor_th}
\begin{align}
    \mathrm{Cov}(s_f^i,s_0^i)
    &= \mathrm{Cov}\!\left(s_0^i2^{-\epsilon}+\frac{\alpha}{\mu}(1+\epsilon)\left(\theta^{i}-r\theta^{i-1}2^{-\epsilon}\right),\,s_0^i\right), \nonumber\\
    &= 2^{-\epsilon}\sigma^2_{s_0} + \frac{\alpha}{\mu}(1+\epsilon)\left[\mathrm{Cov}(\theta^{i},s_0^i)-r2^{-\epsilon}\,\mathrm{Cov}(\theta^{i-1},s_0^i)\right].
    \label{eq:cov_sf_s0_expand_degradation}
\end{align}
Furthermore, $\mathrm{Cov}(\theta^{i},s_0^i) = 0$ and
\begin{eqnarray}
     \mathrm{Cov}(\theta^{i-1},s_0^i) = \frac{1}{2}\,\mathrm{Cov}\!\left(\theta^{i-1},\,s_f^{i-1}\right)
     \nonumber
     = \frac{1}{2}\frac{\alpha}{\mu}(1+\epsilon)\sigma_{\theta}^2,
\end{eqnarray}so that 
\begin{equation}
    1-\zeta_\Delta \equiv \frac{\mathrm{Cov}(s_f, s_0)}{\sigma^2_{s_0}}\;=\; \frac{1}{2^\epsilon}\left[1-(1+\epsilon)^2\frac{r}{2}\Big(\frac{\alpha}{\mu}\Big)^2 \frac{\sigma_{\theta}^2}{\sigma_{s_0}^2}\right].
    \label{eq:zetaG_general_deg}
\end{equation}

From Eq.~(\ref{eq:zetaG_general_deg}), we can compute the division control parameter $\lambda_G$ as
\begin{equation}
    \lambda_G = 1-\frac{1-\zeta_{\Delta}}{2} \;=\; 1-\frac{1}{2^{\epsilon+1}}\left[1-(1+\epsilon)^2\frac{r}{2}\Big(\frac{\alpha}{\mu}\Big)^2 \frac{\sigma_{\theta}^2}{\sigma_{s_0}^2}\right].
    \label{eq:lambdaG_general_deg}
\end{equation}
\subsection*{Size fluctuations}
We compute the variance of cell size at birth using the law of total variance (Eq.~(\ref{eq:sig_total})).

\paragraph*{Calculation of variance for fixed threshold:} The birth-size variance at a fixed threshold has already been calculated in \cite{ElGamel2024}, yielding
\begin{equation}\label{eq:sig_int_deg}
\big\langle \sigma^2_{s_0|\theta}\big\rangle
\;=\;
\Big(\frac{\alpha}{\mu}\Big)^2 (1+\epsilon)^2\frac{(1-r2^{-\epsilon})\langle \theta\rangle}{(4-2^{-2\epsilon})},
\end{equation}
which converges to Eq.~(\ref{eq:sig_int}) for $\epsilon \to 0$.

\paragraph*{Calculation of variance for threshold fluctuations:}
To compute, $\sigma^2_{\langle s_0 \mid \theta\rangle}
\equiv \mathrm{Var}\Big(\,\langle s_0^i \mid \{\theta\}\rangle\Big) $ we use Eq.~(\ref{eq:eq:s_f_degradation}) and the hypothesis of uncorrelated threshold to obtain
\begin{align}
\sigma^2_{\langle s_0 \mid \theta\rangle}
&=\sigma_{\theta}^2\left(\frac{\alpha}{\mu}\right)^2(1+\epsilon)^2\frac{(1-r2^{-2\epsilon}+r^22^{-2\epsilon})}{(4-2^{-2\epsilon})}.
\label{eq:sig_ext_deg}
\end{align}
Now using Eqs.~(\ref{eq:sig_int_deg}) and (\ref{eq:sig_ext_deg}), we obtain the total birth size variance as
\begin{equation}\label{eq:var_total_deg}
    \sigma_{s_0}^2 = \frac{(1+\epsilon)^2}{(4-2^{-2\epsilon})}\left(\frac{\alpha}{\mu}\right)^2\left[(1-r2^{-\epsilon})\,\langle \theta\rangle+\sigma_{\theta}^2(1-r2^{-2\epsilon}+r^22^{-2\epsilon})\right].
\end{equation}

In the limit $\epsilon \to 0$, Eq.~(\ref{eq:zetaG_general_deg}) and Eq.~(\ref{eq:var_total_deg}) reduce to Eq.~(\ref{eq:zetaG_general}) and Eq.~(\ref{eq:var_total}), respectively, with $\rho \to 0$ in the case of uncorrelated thresholds. In the opposite limit, $\epsilon \gg 1$, one finds
$
\lambda_G \simeq
1-\frac{1}{2^{\epsilon+1}}
\left[
1-\frac{2r\sigma_\theta^2}{\langle\theta\rangle+\sigma_\theta^2}
\right]
\to 1,
$ thus driving the system toward increasingly sizer-like behavior. The $\sigma^2_{\theta} \to 0$ limit, $\lambda_G \to 1- 2^{-(\epsilon +1)}$, was also studied in~\cite{ElGamel2024}.

\section{Effects of constant production of divisor protein}
Including a constant production term $\nu$, in the synthesis dynamics of the divisor protein modifies Eq.~(\ref{eq:s_f}), introduced in Section~\ref{sec:div_controll_cor_th}, which now takes the more general form \begin{eqnarray}\label{eq:eq:s_f_prod}
    s_f^i
    &=&s_0^i+\frac{\alpha}{\mu}\left(\theta^{i}-r\theta^{i-1}-\nu\tau^i\right), 
\end{eqnarray}
where $\tau^i$ is the division time at which $x^i(t)$ reaches the threshold level for the first time. In the limit of $\nu \to 0$, the above Eq.~(\ref{eq:eq:s_f_prod}) converges to Eq.~(\ref{eq:s_f}). We will now use Eq.~(\ref{eq:eq:s_f_prod}) to derive the expressions for size fluctuations and division control parameter. For simplicity, we assume $\rho = 0$ in the subsequent calculations.

\subsection*{Linear regression between birth and division size}
Following the same steps as in Section~\ref{sec:div_controll_cor_th}, 
\begin{align}
    \mathrm{Cov}(s_f^i,s_0^i)
    &= \mathrm{Cov}\!\left(s_0^i+\frac{\alpha}{\mu}\left(\theta^{i}-r\theta^{i-1}-\nu\tau^i\right),\,s_0^i\right), \nonumber\\
    &= \sigma^2_{s_0} + \frac{\alpha}{\mu}\left[\mathrm{Cov}(\theta^{i},s_0^i)-r\,\mathrm{Cov}(\theta^{i-1},s_0^i)-\nu\,\mathrm{Cov}(\tau^i,s_0^i)\right].
    \label{eq:cov_sf_s0_expand_prod}
\end{align}
Furthermore, $\mathrm{Cov}(\theta^{i},s_0^i) = 0$. By considering small noise approximation around mean value $s_0^i = \langle s_0\rangle + \Delta_i$, the other two covariances will be given as below:
\begin{eqnarray}
    \text{Cov}(s_0^i,\tau^i) &=&\text{Cov}\left(s_0^i,\frac{1}{\alpha}\ln\left(\frac{s_f^i}{s_0^i}\right)\right)\\
    &=&\frac{1}{\alpha}\left[\text{Cov}\left(s_0^i,\ln\left(1+\frac{\Delta_{i+1}}{\langle s_0\rangle}\right)\right)-\text{Cov}\left(s_0^i,\ln\left(1+\frac{\Delta_{i}}{\langle s_0\rangle}\right)\right)\right]\\
    &\approx&\frac{1}{\alpha}\left[\text{Cov}\left(s_0^i,\frac{\Delta_{i+1}}{\langle s_0\rangle}\right)-\text{Cov}\left(s_0^i,\frac{\Delta_{i}}{\langle s_0\rangle}\right)\right]\\
    &\approx&\frac{\sigma_{s_0}^2}{\alpha\langle s_0\rangle}\left(\frac{1-\zeta_\Delta}{2}-1\right),\quad[1-\zeta_\Delta = \text{Cov}(s_f^i,s_0^i)/\sigma_{s_0}^2]
     \label{eq:cov_1_prod}
\end{eqnarray}

\begin{eqnarray}
    \text{Cov}(s_0^i,\theta^{i-1}) &=&\frac{1}{2}\text{Cov}(s_f^{i-1},\theta^{i-1})
    =\frac{1}{2}\frac{\alpha}{\mu}\text{Cov}(\theta^{i-1},\theta^{i-1})-\frac{\nu}{2}\frac{\alpha}{\mu}\text{Cov}(\tau_{i-1},\theta^{i-1})\\    &=&\frac{\alpha\sigma_{\theta}^2}{2\mu}-\frac{\nu\alpha}{2\mu}\text{Cov}\left(\frac{1}{\alpha}\ln\left(\frac{s_f^i}{s_0^i}\right),\theta^{i-1}\right)\\
    &=&\frac{\alpha\sigma_{\theta}^2}{2\mu}-\frac{\nu}{2\mu\langle s_0\rangle}\text{Cov}\left(\Delta_{i},\theta^{i-1}\right)\\
    \text{Cov}(s_0^i,\theta^{i-1}) &=& \frac{\alpha\sigma_{\theta}^2}{\mu(2+\omega)},
    \label{eq:cov_2_prod}
\end{eqnarray}
where $\omega = \nu/(\mu\langle s_0\rangle)$. Now, substituting Eqs.~(\ref{eq:cov_1_prod}) and (\ref{eq:cov_2_prod}) into Eq.~(\ref{eq:cov_sf_s0_expand_prod}), we obtain
\begin{equation}
    1-\zeta_\Delta = \frac{\mathrm{Cov}(s_f, s_0)}{\sigma^2_{s_0}} = \frac{1}{1+\omega/2}\left(1+\omega-\frac{r}{2+\omega}\left(\frac{\alpha}{\mu}\right)^2\frac{\sigma_{\theta}^2}{\sigma_{s_0}^2}\right).
 \label{eq:zetaG_general_prod}
\end{equation} From Eq.~(\ref{eq:zetaG_general_prod}), we can compute the division control parameter $\lambda_G$ as
\begin{equation}
    \lambda_G = 1-\frac{1}{2+\omega}\left(1+\omega-\frac{r}{2+\omega}\left(\frac{\alpha}{\mu}\right)^2\frac{\sigma_{\theta}^2}{\sigma_{s_0}^2}\right).
 \label{eq:lambdaG_general_prod}
\end{equation}

\subsection*{Size fluctuations}
We compute the variance of cell size at birth using the law of total variance (Eq.~(\ref{eq:sig_total})).

\paragraph*{Calculation of variance for fixed threshold.} The birth-size variance at a fixed threshold has already been calculated in \cite{ElGamel2024}, yielding
\begin{equation}\label{eq:sig_int_prod}
\big\langle \sigma^2_{s_0|\theta}\big\rangle
\;=\;
\left(\frac{\alpha}{\mu}\right)^2\left(\langle \theta\rangle(1-r)-\omega\langle s_0\rangle\frac{\mu}{\alpha}\ln 2\right)\frac{1}{(3+2\omega)},
\end{equation}
which converges to Eq.~(\ref{eq:sig_int}) for $\omega \to 0$.

\paragraph*{Calculation of variance for threshold fluctuations.} To compute $
\sigma^2_{\langle s_0 \mid \theta\rangle}
\equiv \mathrm{Var}\Big(\,\langle s_0^i \mid \{\theta\}\rangle\Big), $
we use Eq.~(\ref{eq:eq:s_f_prod}) and the hypothesis of uncorrelated threshold to obtain
\begin{align}
\sigma^2_{\langle s_0 \mid \theta\rangle}
&=\sigma_{\theta}^2\left(\frac{\alpha}{\mu}\right)^2\,\,\frac{1 + r^2 - \frac{2(r + \omega)}{2 + \omega} + \frac{\omega(\omega + 1)}{1 + \omega/2} \cdot \frac{r}{2 + \omega}}{3 - \frac{\omega(\omega + 1)}{1 + \omega/2}}.
\label{eq:sig_ext_prod}
\end{align}
Now using Eq.~(\ref{eq:sig_int_prod}) and Eq.~(\ref{eq:sig_ext_prod}), we can write the total birth size variance as
\begin{equation}\label{eq:var_total_prod}
    \sigma_{s_0}^2 = \left(\frac{\alpha}{\mu}\right)^2\,\,\left[\left(\langle \theta\rangle(1-r)-\omega\langle s_0\rangle\frac{\mu}{\alpha}\ln 2\right)\frac{1}{(3+2\omega)}+\sigma_{\theta}^2\frac{1 + r^2 - \frac{2(r + \omega)}{2 + \omega} + \frac{\omega(\omega + 1)}{1 + \omega/2} \cdot \frac{r}{2 + \omega}}{3 - \frac{\omega(\omega + 1)}{1 + \omega/2}}\right].
\end{equation}

In the limit of $\omega \to 0$, Eq.~(\ref{eq:zetaG_general_prod}) and Eq.~(\ref{eq:var_total_prod}) converge to Eq.~(\ref{eq:zetaG_general}) and Eq.~(\ref{eq:var_total}), respectively, with $\rho \to 0$ for uncorrelated threshold. For $\omega \gg 1$, instead, $\lambda_G
\simeq
\frac{1}{2+\omega}
+
\frac{2r\ln 2\,\sigma_{\theta}^2}
{
(1-r)\langle \theta\rangle
+
\omega \ln 2 \,(1-2r-r^2)\sigma_{\theta}^2
} \propto \frac{1}{\omega} \to 0$ at fixed  $\sigma^2_\theta$, thus driving the system towards increasingly timer-like behavior. The $\sigma^2_\theta \to 0$ limit, $\lambda_G = 1/(2+\omega)$, was also studied in~\cite{ElGamel2024}.

\begin{table}[t]  
	\begin{ruledtabular} 
\begin{tabular}{rl} Symbol & Description \\ \hline $s$ & Cell size \\ $x$ & Copy number of divisor protein \\ $s_0^i$ & Birth size in cycle $i$ \\ $s_f^i$ & Division size in cycle $i$ \\ $r$ & Protein reset fraction \\ $\tau$ & Division time \\ $\alpha$ & Exponential growth rate \\ $\mu$ & Size-dependent production rate of $x$ \\ $\nu$ & Constant production rate of $x$ \\ $\gamma$ & Degradation rate of $x$ \\ $\theta$ & Threshold level \\ $\rho$ & Threshold memory between cycles \\ $\eta$ & Extrinsic threshold noise \\ $\sigma_{\rm extr.}^2$ & Variance of extrinsic noise \\ $\langle \cdot \rangle$, $\mathbb{E}[\cdot]$ & Ensemble mean \\ $\sigma^2_{(\cdot)}$, $\mathrm{Var}[\cdot]$ & Ensemble variance \\ CV & Coefficient of variation \\ $1-\zeta_\Delta$ & Birth--division regression slope \\ $\lambda_G$ & Division-control parameter \\ $\Sigma$ & Relative extrinsic/molecular noise strength \\ \end{tabular} 
\end{ruledtabular} 
\caption{\label{tab:notation} Symbols used throughout the manuscript.}
\end{table}

\begin{figure}[ht!]
    \centering
\includegraphics[scale=0.73]{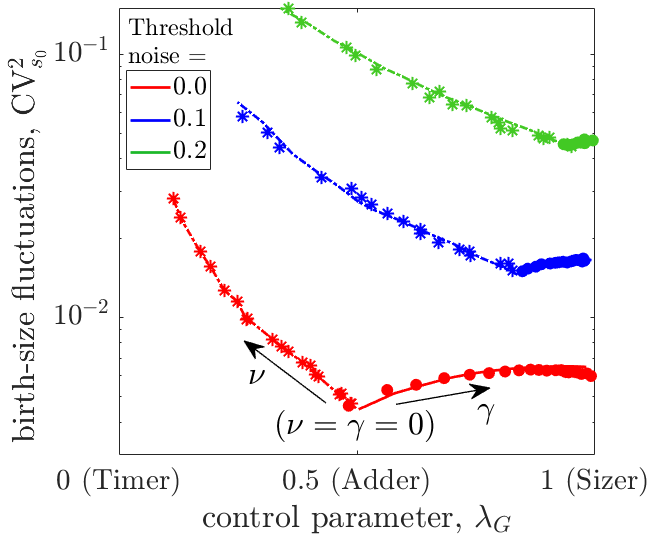}
    \caption{\textbf{The noise in birth size is optimum when the degradation rate ($\gamma$) and the constant production rate ($\nu$) of divisor protein are zero. The optimum value shifts from adder ($\lambda_G = 0.5$) to sizer ($\lambda_G = 1$) with threshold noise.} In the plot, the symbols are obtained from the simulation, and the curves are obtained from the analytical expressions. The parameter values are: $\langle \theta\rangle$ = 150, $\mu$ = 1, $\alpha$ = 0.01, $r$ = 1/2, $\rho=0$. For the solid lines, $\nu = 0$ and $\gamma$ increases from 0 to 0.01. We plot Eq.~(\ref{eq:lambdaG_general_deg}) as x-axis and  y-axis is (\ref{eq:var_total_deg}) normalized by $\langle s_0\rangle^2$. For the dashed lines, $\gamma = 0$ and $\nu$ increases from 0 to 1, and we plot Eq.~(\ref{eq:lambdaG_general_prod}) as x-axis. The y-axis is Eq. (\ref{eq:var_total_prod}) normalized by $\langle s_0\rangle^2$. The optimum in the birth size fluctuations is obtained when $\gamma = \nu = 0$.}
    \label{fig:placeholder}
\end{figure}

\begin{figure}[htpb]
    \centering
\includegraphics[scale=1]{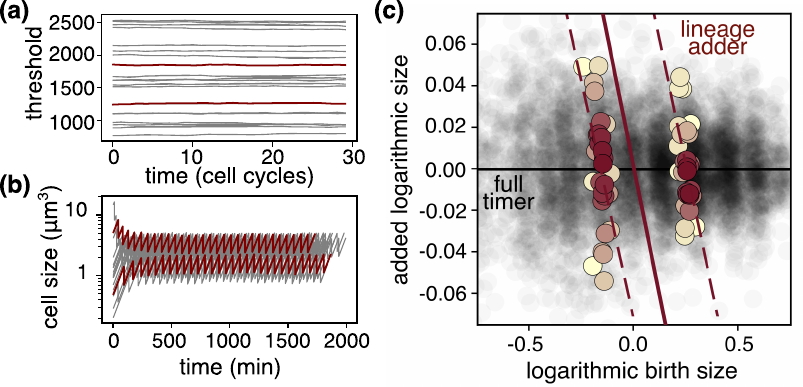}
    \caption{\textbf{Timer control emerges from averaging lineage-specific adders in the fully autocorrelated limit.} (a)-(b) Examples of simulated threshold and cell size dynamics in the $\rho \to 1$ limit, showing stratified threshold levels arising from lineage-dependent initial conditions. (c) Size–growth plot illustrating a Simpson’s paradox–like effect, where population-level timer behavior emerges from individual lineage-specific adders. Simulations with parameters: $\theta = 1000, \Sigma = 0.1, \rho = 0.99999, r=0.5, \alpha = 0.0116 \ \text{min}^{-1}, \mu = 5.8 \ \text{min}^{-1} \mu m^{-3}$.}
\end{figure}

\clearpage
\bibliography{references}